\documentclass[times,twocolumn]{aastex631}

\shorttitle{X-ray shots in Cyg~X-1}
\shortauthors{Qin, Feng \& Tao}

\graphicspath{{./}{fig/}}

\newcommand\SI[2]{#1~\mathrm{#2}}
\newcommand\qtyrange[3]{{#1\textendash#2~\mathrm{#3}}}
\newcommand\subrm[1]{_{\mathrm{#1}}}

\newcommand\delchi{\Delta\chi^2}
\newcommand\delchith{\delchi\subrm{th}}

\usepackage{txfonts}

\hypersetup{%
  pdfstartview      = FitH,
  bookmarksnumbered = true,
  bookmarksopen     = true,
  pdftitle          = {X-ray shots in Cyg X-1},
  pdfauthor         = {Qin, Feng \& Tao},
}

\begin{document}

\title{A new algorithm for detecting X-ray shots in Cyg~X-1}

\author{Jin Qin}
\affiliation{Department of Astronomy, Tsinghua University, Beijing 100084, China}

\author[0000-0001-7584-6236]{Hua Feng}
\email{hfeng@ihep.ac.cn}
\affiliation{Key Laboratory of Particle Astrophysics, Institute of High Energy Physics, Chinese Academy of Sciences, Beijing 100049, China}

\author{Lian Tao}
\affiliation{Key Laboratory of Particle Astrophysics, Institute of High Energy Physics, Chinese Academy of Sciences, Beijing 100049, China}

\begin{abstract}
The short-term X-ray variability of Cyg~X-1 can be interpreted as random occurrence of mini-flares known as the shots, whose physical nature is still unclear. 
We propose a new algorithm for shot identification in the X-ray light curve, based on baseline detection and template fitting. 
Compared with previous techniques, our algorithm allows us to detect shots with lower amplitudes and shorter time separations. 
With \textit{NICER} observations, we find that, after correction for detection sensitivity, both the shot amplitude and recurrence rate are positively scaled with the mean count rate, while the recurrence rate has a much higher dependence on the count rate. 
These suggest that a higher mass accretion rate will drive more and slightly larger shots. 
We also find that the abrupt hardening near the shot peak found in previous studies is attributed to different shot profiles in different energy bands; there is no need to involve a rapid physical process to suddenly harden the emitting spectrum.  

\end{abstract}

\keywords{}

\section{Introduction}\label{sec:intro}

The Galactic X-ray binary Cygnus~X-1 (Cyg~X-1) consists of a black hole with a mass of $21.2 \pm 2.2 \, M_\sun$ and a high-mass companion star. 
The binary system has an orbital period of $\SI{5.6}{d}$ and an inclination of $27\fdg5^{+0\fdg8}_{-0\fdg6}$, and is located at a distance of $2.2^{+0.2}_{-0.2}\,\mathrm{kpc}$ \citep{2021Sci...371.1046M}. The X-ray emission is persistent and transitions between two major spectral states, the low/hard and high/soft state \citep{1972ApJ...177L...5T}.

Cyg~X-1 shows strong X-ray variability over a variety of timescales from milliseconds to years,
among which the non-periodic variability known as ``shots'' resembles second-duration mini-flares \citep{1972ApJ...174L..35T}.
The shots have been observed in the full X-ray band, from $\sim$$\SI{0.1}{keV}$ up to $\sim$$\SI{200}{keV}$ \citep{2013ApJ...767L..34Y,2022MNRAS.512.6067B}, and also in different spectral states \citep{1999ApJ...514..373F}. 
Similar shots have also been observed in other black hole X-ray binaries, e.g., GX~339--4, GS~2023+338, and GS~1124--68 \citep{1999NuPhS..69..344N}. 
Since the X-ray is emitted in the immediate proximity to the black hole, from the inner accretion disk, corona, or jet base, studying the shots can help us further understand the physical processes of the hot plasma around the black hole.

Shots in Cyg~X-1 have been intensively studied in the past. 
\citet{1972ApJ...174L..35T} found that the Cyg~X-1 shots were consistent with random pulses that resembled the observed power spectra. 
While the rough shape of power spectrum can be reproduced by the shot model with appropriate shot parameters \citep[e.g.,][]{1989Natur.342..773M,1990A&A...227L..33B,1991ApJ...376..295L}, it is difficult to explain other details and the amplitude variations of the power spectrum, e.g., the rms-flux relation \citep{2001MNRAS.323L..26U}. 
The average shot profile was revealed to be slightly asymmetric with a slow rise and a fast decay \citep{1989Natur.342..773M,1999ApJ...514..373F,2001ApJ...554..528N}. 
The profile can be modeled by the sum of two exponential components, a fast one with a timescale of $\sim$$\SI{0.1}{s}$ and a slow one with a timescale of $\sim$$\SI{1}{s}$ \citep{1994ApJ...423L.127N,1999ApJ...514..373F,2001ApJ...554..528N}. 
The shot profile was found to be narrower and more asymmetric in the higher energy band, and vary in different spectral states \citep{1999ApJ...514..373F}. 
Both the shot amplitude and waiting time (time between shots) follow an exponential distribution \citep{1995ApJ...452L..49N}; whether or not there is a deficit of shot occurrence at short time separations is controversial \citep{1995ApJ...452L..49N,2005ApJ...633.1085F}. 
\citet{2002PASJ...54L..69N} further suggested that the two distributions could be log-normal, implying a physical link between shots and gamma-ray bursts due to similar behaviors \citep{1996ApJ...469L.115L}. 

The hardness or spectral parameters were found to vary with time within a shot \citep{1994ApJ...423L.127N,1999ApJ...514..373F,2013ApJ...767L..34Y,2022MNRAS.512.6067B}. In particular in the hard state, the hardness ratio gradually decreases before the peak but abruptly returns to the average around the peak \citep{1994ApJ...423L.127N,1999ApJ...514..373F,2013ApJ...767L..34Y}. 
Below the frequency of a few Hz, the Fourier power and cross spectra, at least the shapes, can be interpreted as due to shot variability \citep{2001ApJ...554..528N}.
A near-unity coherence between different \textit{RXTE} energy bands was obtained in this frequency range \citep{1999ApJ...510..874N,2001ApJ...554..528N}. 
Also, a hard time lag was found \citep{1988Natur.336..450M,1999ApJ...514..373F}.

Based on these observations, several possible physical mechanisms for the shot have been proposed. 
The magnetic flare model suggests that the energy release of magnetic reconnections on the accretion disk is responsible for the shots \citep{1979ApJ...229..318G,1982MNRAS.198..689P,1999MNRAS.306L..31P}, and is able to explain the prompt spectral hardening \citep{2013ApJ...767L..34Y}. 
The disturbance propagation is another physical model to explain the shot behavior \citep{1996ApJ...464L.135M}.
With numerical simulations, \citet{1996ApJ...464L.135M} demonstrated that the propagation and reflection of the disturbance wave can naturally reproduce the observed shot profile. 
When the disturbance propagates toward the innermost region where the radius is comparable to the characteristic scale of the disturbance, the disturbance will evolve from the initial thermal mode into a mixture of thermal and acoustic modes because the plasma is inhomogeneous. 
While the inward modes are swallowed by the black hole, the acoustic mode can propagate outward, producing the decaying tail of the shot \citep{1996PASJ...48...67K,1996ApJ...464L.135M}.
Although the disturbance propagation model can successfully reproduce the shot profile, it has difficulties in explaining the prompt hardening around the peak.
On the other hand, the magnetic flare model can reproduce the prompt hardening but is difficult to explain the gradual hardness decrease before the peak.
Therefore, some authors proposed that both mechanisms might have been involved \citep{2001ApJ...554..528N,2013ApJ...767L..34Y}. 

In the previous studies, shots were detected by techniques based on their relatively high fluxes \citep[e.g.,][]{1994ApJ...423L.127N,2005ApJ...633.1085F,2022MNRAS.512.6067B}. 
These techniques perform well for strong shots but are less efficient in detecting weak ones. 
Moreover, they may have difficulties in distinguishing proximate shots and lead to biases in some statistical results \citep[see][]{2005ApJ...633.1085F}. 
In this paper, we propose a new technique that is unbiased for shots with small amplitudes and temporal proximity, enabling us to obtain a more complete and unbiased shot sample. 
We describe the observations and introduce our new shot detection algorithm in \autoref{sec:obs-tech}, including comparisons with previous techniques, and present the results in \autoref{sec:shot-ana}. The shot mechanisms are discussed in \autoref{sec:disc}. 

\section{Observations and Detection Algorithm}\label{sec:obs-tech}

Neutron Star Interior Composition Explorer \citep[\textit{NICER},][]{2012SPIE.8443E..13G} observations of Cyg~X-1 with an effective exposure ${>\SI{200}{s}}$ as of 2023 January 1 are adopted.
The level~2 data are downloaded from HEASARC and reprocessed with the \texttt{nicerl2} tool in \texttt{HEASoft} 6.33.2 with the \textit{NICER} CALDB version 20240206.
We extracted the $\qtyrange{0.25}{12}{keV}$ light curves with a time resolution of $\SI{0.05}{s}$ using the tool \texttt{nicerl3-lc}, normalized to 52 FPMs.
Cyg~X-1 may show dips in its light curve, possibly due to absorption of clumpy winds or clumps in the accretion flow \citep{2002ApJ...564..953F}.
We identified such dips with a duration of $\sim$$\SI{1}{s}$ in 13 observations\footnote{ObsIDs 1100320101, 1100320102, 1100320103, 1100320104, 1100320106, 1100320107, 1100320119, 1100320120, 1100320121, 1100320122, 2636010102, 5100320129, and 5100320134.} and discarded them, as
dips with a duration similar to that of the shot may cause problems in shot detection. 
We also found that when the count rate is high, the shots occur so frequently that the typical waiting time is similar to the shot duration, which may affect the shot extraction.
We therefore excluded another 17 observations\footnote{ObsIDs 0100320102, 0100320110, 1100320108, 1100320109, 1100320118, 2636010201, 5100320120, 5100320124, 5100320125, 5100320126, 5100320127, 5100320128, 5100320130, 5100320131, 5100320132, 5100320139, and 5100320140.} and only used those with an average count rate below $\SI{12000}{counts~s^{-1}}$.
In summary, 30 observations are discarded and 70 observations with a total exposure of 402~ks are used in this work. 

We fit the $\qtyrange{3}{12}{keV}$ \textit{NICER} spectra with a single power-law model to determine the spectral state, and found that the photon index is always less than 2.2 in all observations, corresponding to the hard or intermediate state, but not the soft state \citep{2013A&A...554A..88G}. 
Following \citet{2013A&A...554A..88G}, we also checked the MAXI \citep{2009PASJ...61..999M} and \textit{Swift}/BAT \citep{2013ApJS..209...14K} data to cross-check the spectral state of Cyg~X-1 during the \textit{NICER} observations, and obtained consistent results. 

The new shot detection technique is based on the baseline detection and template fitting.  
Shots are identified from a baseline-subtracted light curve by fitting with an average shot profile iteratively. The light curve is in the unit of counts.

(a) Baseline subtraction. 
The baseline $b_i$ is subtracted from the original light curve $c_i$ to get the new one ${n_i=c_i-b_i}$, in which the long-term variations are removed. 
We adopted the Asymmetric Least Squares algorithm \citep[\texttt{asls};][]{asls} for baseline detection. 
This technique calculates the baseline by minimizing the object function
\begin{equation}\label{eq:asls}
\sum_{i=1}^N w_i(c_i-b_i)^2 + \lambda\sum_{i=1}^{N-d}{(\Delta^d b_i)^2} \, , \quad 
w_i=\left\{\begin{array}{ll}
  p, & c_i> b_i \\
  1-p, & c_i\le b_i
\end{array}\right. \, ,
\end{equation}
where $N$ is the number of bins in the light curves, $\Delta^d$ is the $d$th order finite-difference operator, and $\lambda$ is a penalty scale factor. 
The first term in \autoref{eq:asls} represents the contribution from residuals, and a small $p$ ($<0.5$) tends to generate positive residuals. 
The second term is associated with the roughness, which requires the baselines to vary smoothly.
The baseline is calculated using the Python library \texttt{pybaselines} \citep{pybaselines}, with ${\lambda=100}, {p=0.05}$ and ${d=2}$.

(b) Profile construction. 
An average shot profile is constructed using strong and isolated shots identified from the baseline-subtracted light curve. 
A time bin is marked as the potential peak position of a shot if it has the maximum count over the neighboring 21 bins ($\SI{1.05}{s}$). 
Initially, shots with a peak count greater than a certain threshold $n\subrm{th}$ are selected and summed to generate the shot profile by aligning the peak position. The shot profile is normalized to have a unity peak amplitude.
To exclude superimposed shots, we fit each individual shot with the average profile,
\begin{equation}\label{eq:chi2-fit}
\chi^2 = \sum_{i=-10}^{10} \frac{(n_i-p_iA)^2}{\sigma^2(c_i)} \, ,
\end{equation}
where $n_i$ and $p_i$ are the counts of the data and normalized shot profile at bin $i$, respectively, $A$ is the baseline-subtracted peak amplitude of the shot, and $\sigma^2(c_i)$ is the variance of the data. 
Shots with ${\chi^2/20<C}$ are used to construct a new profile, where $C$ is an empirical cut for the goodness of fit. 
This procedure is repeated until the input and output shot samples become identical.

(c) Shot detection. 
The shots are detected by slide fitting with the shot profile in the light curve one after another. 
By moving the average shot profile through the whole light curve, one fits the shot amplitude at each time bin and finds the strongest shot that leads to the largest $\delchi$. A shot component is then added into the model light curve and the search is repeated to find the next strongest one until $\delchi$ due to the new shot component is lower than a given threshold $\delchith$.

\begin{figure}
\centering
\includegraphics[width=\columnwidth]{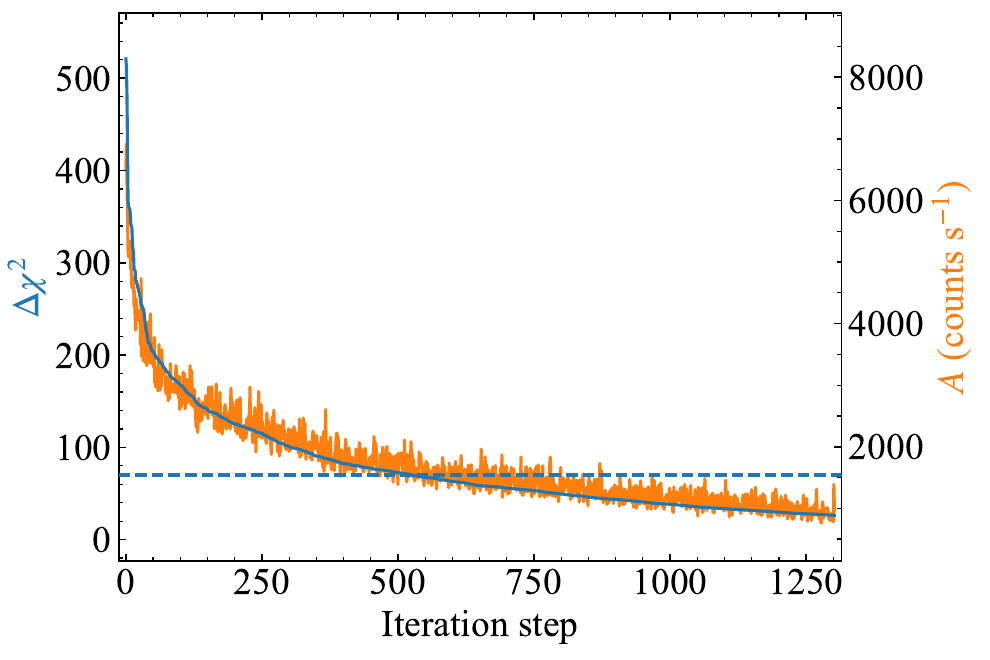}
\caption{$\delchi$ (blue solid) and shot peak amplitude $A$ (orange) as a function of the iteration step. The 8th GTI in the \textit{NICER} observation 5100320101 with an exposure of $\SI{1.148}{ks}$ is used as an example. The blue dashed line marks $\delchith=70$.}
\label{fig:chi2curve}
\end{figure}

When applying the technique on the \textit{NICER} data, in each observation, we chose an $n\subrm{th}$ such that the stronger half of shots are selected. 
We set $C = 5$ in step~(b).
Different $n\subrm{th}$ and $C$, which only determine the shot profile, have almost no influence on the scientific results.
The stopping criterion is set as $\delchith = 70$. 
\autoref{fig:chi2curve} shows $\delchi$ and the shot amplitude $A$ vary as a function of the iteration step. 
As one can see, $\delchi$ is roughly scaled with the amplitude of detected shots.
We obtained consistent results if $\delchith = 60$ is used.

\begin{figure}
\centering
\includegraphics[width=.7\columnwidth]{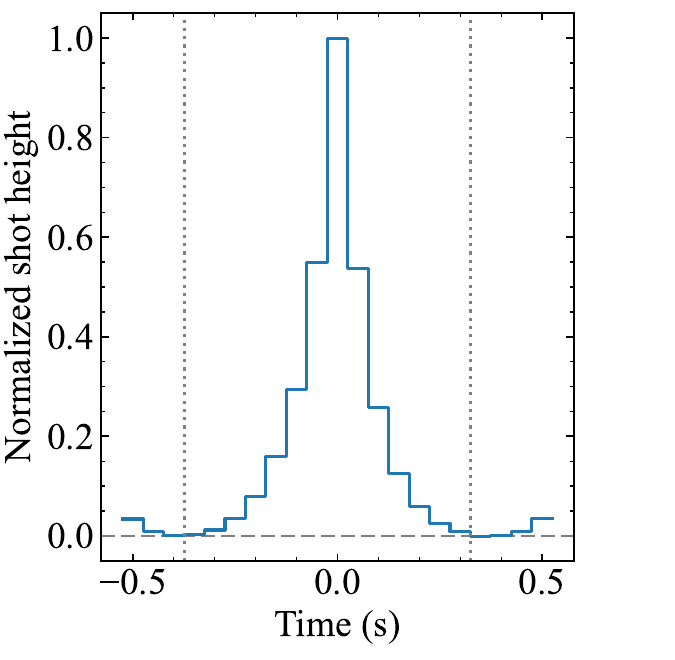}
\caption{Normalized shot profile generated using the \textit{NICER} data.  The part between the vertical dotted lines is used for light curve fitting in step (c). The horizontal dashed line marks the zero level in the baseline-subtracted light curve. The errors are smaller than the line width.}
\label{fig:profile}
\end{figure}

\begin{figure*}
\centering
\includegraphics[width=\linewidth]{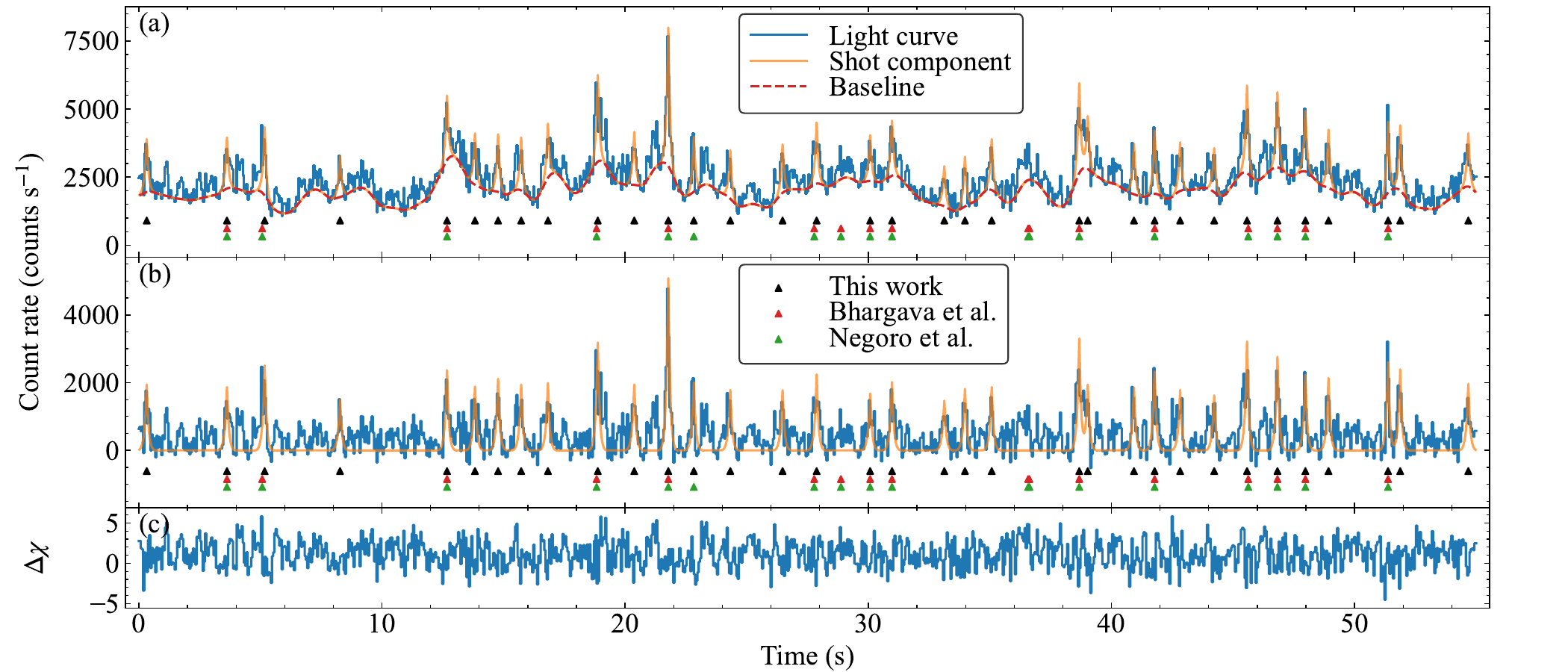}
\caption{\textit{NICER} light curve to illustrate the shot detection algorithm: (a)~the original light curve with the best-fit baseline and shot components, (b)~baseline-subtracted light curve with shot components, and (c)~residuals. {The triangles mark the shots detected using algorithms in \citet{1994ApJ...423L.127N}, \citet{2022MNRAS.512.6067B}, and this work.}}
\label{fig:nicer-fit-res}
\end{figure*}

We examined the shot profiles generated in different observations and found that they have a small variation. 
They have almost the same normalized shape, although the amplitudes may vary.
Thus, a combined shot profile from all observations is used, shown in \autoref{fig:profile}. 
We notice that there is convex (deficit in count rate) on both wings. This is simply because we have used isolated shots to construct the profile. Shots outside the time window raise the outer wings. If we do not require isolated shots for the profile, the composite shot profile decreases monotonically away from the peak, just like those shown in previous studies. In step (c), we only use the central part of the shot profile (see \autoref{fig:profile}), and the convex has no impact on the results. 
We note that our shot profile shows only a single temporal component with an exponential timescale of 0.07--0.08~s (fitted with an exponential function). 
It is hard to determine whether or not there exists a second, slow component ($\sim$1~s) as reported in the literature \citep{1994ApJ...423L.127N,1999ApJ...514..373F,2001ApJ...554..528N}, due to shot superpositions.

With this algorithm, we successfully detected shots in the 70 \textit{NICER} observations. 
A segment of \textit{NICER} light curve is displayed in \autoref{fig:nicer-fit-res} for illustration. 

Previous shot detection techniques are mainly based on flux comparison \citep[e.g.,][]{1994ApJ...423L.127N,1999ApJ...514..373F,2013ApJ...767L..34Y,2022MNRAS.512.6067B}. 
For example, they required that the shot peak should have the maximum counts among a certain time interval, and also be significantly higher than the local average \citep{1994ApJ...423L.127N, 2013ApJ...767L..34Y, 2022MNRAS.512.6067B}. 
In \autoref{fig:nicer-fit-res}, we also marked shots detected using the technique described in \citet{1994ApJ...423L.127N} and \citet{2022MNRAS.512.6067B}. 
As one can see, the previous techniques cannot effectively detect shots with small amplitudes and those superimposed on each other, e.g., the shots near 39~s and 52~s in \autoref{fig:nicer-fit-res}.

\section{Shot Analysis}\label{sec:shot-ana}

\subsection{Statistical distributions}\label{sec:stat-ana}

\begin{figure}
\centering
\includegraphics[width=\columnwidth]{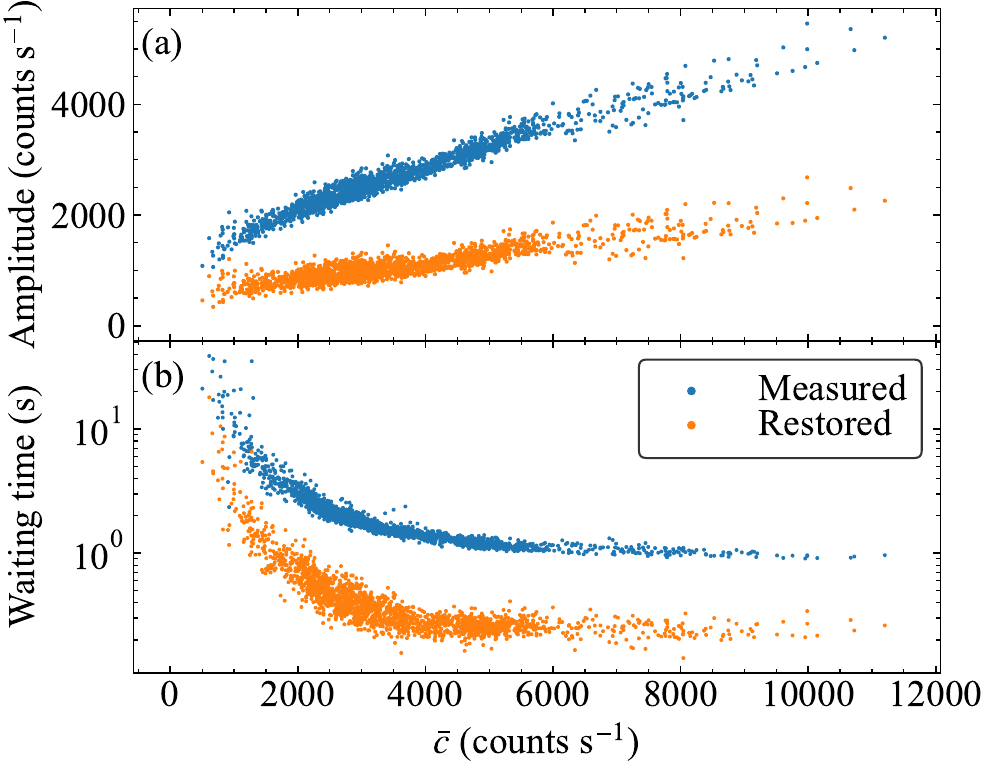}
\caption{Shot amplitude and waiting time versus the count rate $\bar{c}$ averaged over each 200~s segment (blue points). The orange points represent the sensitivity-corrected data.}
\label{fig:seg-statistics}
\end{figure}

\begin{figure}
\centering
\includegraphics[width=\columnwidth]{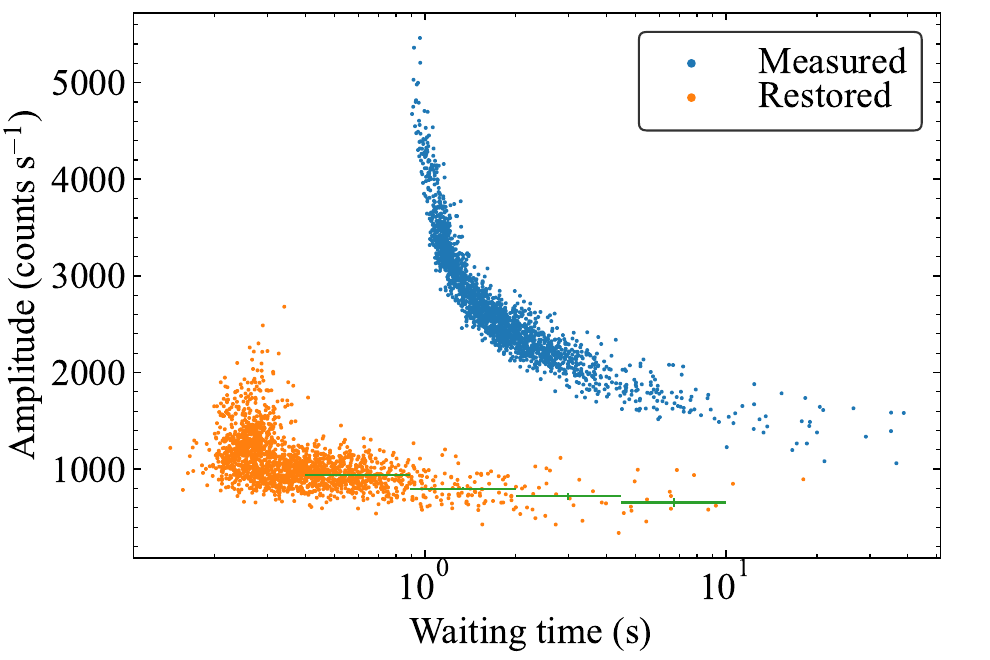}
\caption{Measured and restored relations of shot amplitude vs.\ waiting time averaged over each 200~s segment. The green bars indicate the median $\overline{A_0}$ with $1\sigma$ errors as a function of $\overline{\Delta t_0}$ in the range $\overline{\Delta t_0} \ge 0.4$~s, where the waiting time measurement is not affected by the detection sensitivity.}
\label{fig:amp-wt-relation}
\end{figure}

We divided the light curves into segments with a duration of 200~s, and for each segment, calculated the average count rate $\bar{c}$, the average shot amplitude $\bar{A}$, and the average waiting time $\overline{\Delta t}$, defined as the time interval between successive shots. 
We plotted $\bar{A}$ and $\overline{\Delta t}$ as a function of $\bar{c}$ in \autoref{fig:seg-statistics}, and 
$\bar{A}$ vs.\ $\overline{\Delta t}$ in \autoref{fig:amp-wt-relation}. 

\begin{figure}
\centering
\includegraphics[width=\columnwidth]{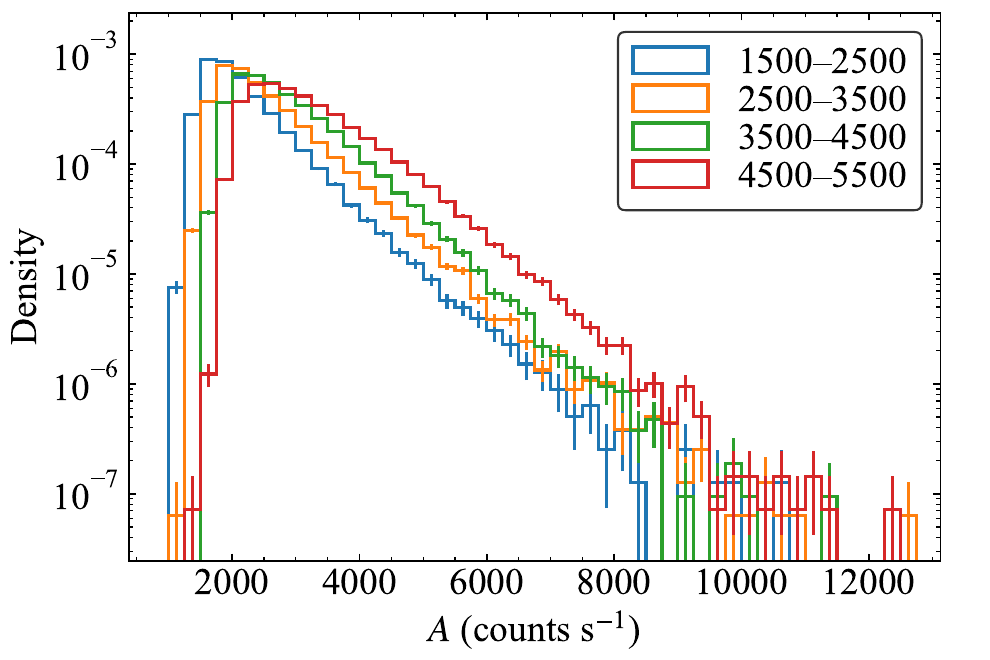}
\caption{Distributions of the $\qtyrange{0.25}{12}{keV}$ shot amplitude in different ranges of mean count rate $\bar{c}$ (in counts~s$^{-1}$, see legend).}
\label{fig:amp-hist}
\end{figure}

The detected sample is incomplete, because the search cannot find relatively weak shots. 
As a consequence, the average amplitude and waiting time measured against the observed sample will be overestimated. 
Corrections are needed to remove the bias.
The intrinsic values can be restored as follows.
When a new shot component is successfully detected (one has $n_i \approx p_i A$), the decrease in $\chi^2$ can be expressed as 
\begin{equation}
\delchi = \sum\frac{n_i^2 - (n_i-p_iA)^2}{\sigma^2(c_i)} 
\approx \sum\frac{p_i^2 A^2}{\bar c}
\propto A^2 / {\bar c} \,,
\end{equation}
suggesting that the sensitivity of shot detection $A_{\min} \propto \Delta \chi \sqrt{\bar{c}}$. 
Previous studies \citep[e.g.,][]{1995ApJ...452L..49N} show that the shot amplitude obeys an exponential distribution, which is also consistent with our findings (\autoref{fig:amp-hist}). 
We assume that the shot amplitude follows an exponential distribution,
\begin{equation}
f(A)= \frac{1}{\overline{A_0}} e^{-A/\overline{A_0}} \, ,
\end{equation}
where $\overline{A_0}$ is the true average amplitude of shots. 
The measured average amplitude is therefore
\begin{equation}\label{eq:amp-cali}
\bar{A} = \frac{
  \int_{A_{\rm min}}^{+\infty} Af(A) \,d A
}{
  \int_{A_{\rm min}}^{+\infty} f(A) \,d A
} = \overline{A_0} + A_{\rm min} \, .
\end{equation}
The true average waiting time $\overline{\Delta{t_0}}$ is related to the measured average waiting time $\overline{\Delta{t}}$ as
\begin{equation}\label{eq:wt-cali}
\frac{\overline{\Delta{t_0}}}{\overline{\Delta{t}}} = \frac{N}{N_0} = \frac{
  \int_{A_{\rm min}}^{+\infty} f(A) \,d A
}{
  \int_{0}^{+\infty} f(A) \,d A
} = e^{-A_{\rm min}/\overline{A_0}},
\end{equation}
where $N$ is the number of detected shots, and $N_0$ is the true number of shots. 
According to Equations~(\ref{eq:amp-cali}) and (\ref{eq:wt-cali}), the intrinsic $\overline{A_0}$ and $\overline{\Delta{t_0}}$ are restored, and plotted as orange points in \autoref{fig:seg-statistics} and \autoref{fig:amp-wt-relation}.
The shot amplitude is positively scaled with the mean count rate. 
The shot waiting time is inversely scaled with the mean count rate until $\overline{\Delta{t}}$ approaches 1~s, which is roughly the detection limit of the shot waiting time. 
The limiting $\overline{\Delta{t}}$ also causes a break in the relations shown in \autoref{fig:amp-wt-relation} at small $\overline{\Delta{t}}$ or $\overline{\Delta{t}_0}$; only the orange data points with $\overline{\Delta{t_0}} \gtrsim 0.4$~s can be regarded as being successfully restored.
We further plot the median $\overline{A_0}$ in different $\overline{\Delta{t}_0}$ ranges, and there is a weak dependence between them.
We tested with other segment durations, e.g., $\SI{50}{s}$ and $\SI{400}{s}$, and the conclusions remain. 

\begin{figure}
\centering
\includegraphics[width=\columnwidth]{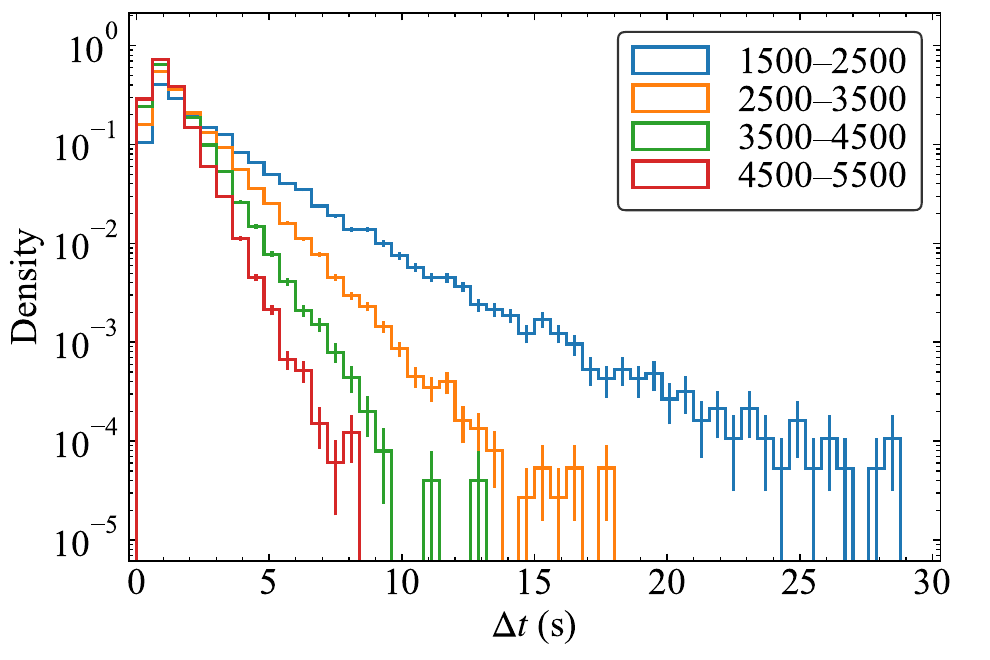}
\caption{Distributions of the waiting time for shots in different ranges of mean count rate $\bar{c}$ (in counts~s$^{-1}$, see legend).}
\label{fig:t12-hist}
\end{figure}

The distributions of the shot amplitude and waiting time are shown in \autoref{fig:amp-hist} and \autoref{fig:t12-hist}, respectively. As the shot properties are tightly correlated with the average count rate, the distributions are divided into sub-groups based on the mean count rate. 
Both the amplitude and waiting time distributions follow an exponential law, with a cutoff below the peak due to detection sensitivity.  
For the waiting time distribution, the exponential slope increases with increasing mean count rate, suggesting that the shot recurrence rate increases with increasing count rate. 

\begin{figure}
\centering
\includegraphics[width=\columnwidth]{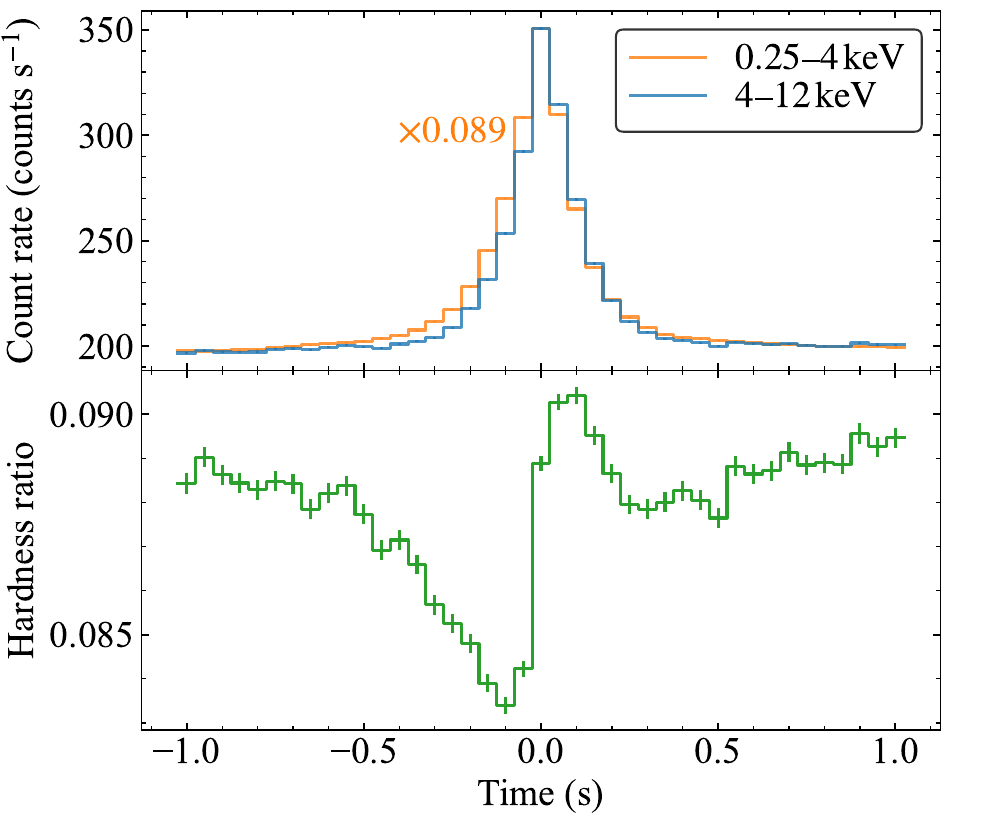}
\caption{Average shot profiles in different energy bands (top) and the corresponding hardness ratio (bottom). The soft band profile is rescaled by multiplying a factor of 0.089 to have the same peak amplitude as the hard one.  The errors of the shot profiles are smaller than the line width.  The hardness ratios are calculated with the unscaled shot profiles.}
\label{fig:hr}
\end{figure}

\subsection{Hardness evolution}\label{sec:hr}

To reproduce the abrupt hardening around the shot peak found in previous studies \citep[e.g.,][]{1994ApJ...423L.127N, 1999ApJ...514..373F, 2013ApJ...767L..34Y}, we plot the hardness ratio variation between two energy bands, $\qtyrange{4}{12}{keV}$ and $\qtyrange{0.25}{4}{keV}$, along with the composite shot profiles in the two bands (see \autoref{fig:hr}).

As one can see, if the two curves are normalized to have the same peak amplitude, the hard band shows a rise pattern narrower than the soft band, leading to a gradual softening before the peak.
The decay timescales in the two bands are almost identical, corresponding to a constant hardness ratio. Thus, there must be an abrupt jump of the hardness near the peak. 

\section{Discussion}\label{sec:disc}

In this paper, we introduced a new algorithm for shot detection. How the detection sensitivity affects the results and how to restore the intrinsic shot amplitude and waiting time is presented in \autoref{sec:stat-ana}. 
Our technique allows us to find shots with a time separation down to $\sim$1~s, close to the shot duration. 
This is consistent with the minimum $\Delta t$ shown in \autoref{fig:t12-hist}, contradicting the conjecture of a deficit of shots below a time separation of $\sim\qtyrange{5}{10}{s}$ \citep{1995ApJ...452L..49N}.  

As shown in \autoref{fig:amp-hist} and \autoref{fig:t12-hist}, the distributions of shot amplitude and waiting time are both exponential. The lower cutoff is consistent with the sensitivity of the technique. In particular, the lower cutoff in the $A$ distribution in \autoref{fig:amp-hist} is scaled with $\bar{c}$. 
For the waiting time, the exponential slope of the distribution is a strong function of the mean count rate. Therefore, if one plots the distribution in a wide range of count rate, one may see a log-normal distribution as reported in the literature \citep{1995ApJ...452L..49N,2002PASJ...54L..69N}. 

The mean count rate can be regarded as the mean mass accretion rate.
The correlations between the mean count rate and the shot properties may imply that the generation of shots is simply modulated by the accretion rate. 
At a higher accretion rate, both the average shot amplitude and shot recurrence rate are higher (\autoref{fig:seg-statistics}).
However, the dependence of the shot amplitude on the waiting time is weak (\autoref{fig:amp-wt-relation}).
These mean that the mass accretion rate has a higher impact on the shot recurrence rate than on the shot amplitude.
These results can be useful in constraining the shot models.

\citet{2001ApJ...554..528N} and \citet{2013ApJ...767L..34Y} proposed that both magnetic flares and disturbance propagation are needed in the shot generation to account for both the long timescale ($\sim$1~s) flux evolution and short timescale ($\sim$0.1~s or less) hardening, with the disturbance propagation for the longer one and the magnetic reconnection for the shorter.
As we demonstrate here that the slightly different flux rising timescales in the two energy bands can lead to a prompt rise of the hardness ratio, there is no need to invoke any short timescale physical processes. 
An energy dependence of the flux rising timescale is also consistent with the disturbance propagation from the cooler outer disk into the inner hotter disk \citep{1999ApJ...514..373F}.
Therefore, the disturbance propagation alone can explain both the shot profile and hardness evolution.

\begin{acknowledgments}
We thank the anonymous referee and statistic editor for useful comments.
HF acknowledges funding support from the National Natural Science Foundation of China under grants Nos.\ 12025301, 12103027, \& 11821303, and the Strategic Priority Research Program of the Chinese Academy of Sciences.
LT acknowledges funding support from the National Natural Science Foundation of China under grant No.~12122306. 
JQ thanks Wei Yu and Panping Li for their helps with the \textit{NICER} data reduction.
\end{acknowledgments}

\vspace{5mm}
\facilities{\textit{NICER}}

\software{%
  pybaselines \citep{pybaselines}
}

\bibliography{refs}{}
\bibliographystyle{aasjournal}

\end{document}